\documentclass[%
 reprint,
 amsmath,amssymb,
 aps,
prl,
]{revtex4-1}
	\usepackage{graphics}
	\usepackage{epsfig}
	\usepackage{graphicx}
	\usepackage{color}
	\usepackage{comment}
	\usepackage{ulem}
\begin{document}
	\unitlength = 1mm
\title{Hybrid Spin-Orbit Exciton-Magnon Excitations in FePS$_3$}

\author{Ramesh Dhakal}
\affiliation{Department of Physics and Center for Functional Materials, Wake Forest University, Winston-Salem, North Carolina 27109, USA}

\author{Samuel Griffith}
\affiliation{Department of Physics and Center for Functional Materials, Wake Forest University, Winston-Salem, North Carolina 27109, USA}

\author{Stephen M. Winter}
\email{winters@wfu.edu}
\affiliation{Department of Physics and Center for Functional Materials, Wake Forest University, Winston-Salem, North Carolina 27109, USA}
\date{\today}

\newcommand\sw[1]{\textcolor{blue}{(SW: #1)}}

\begin{abstract}
FePS$_3$ is a layered van der Waals (vdW) Ising antiferromagnet that has recently been studied in the context of true 2D magnetism, and emerged as an ideal material platform for investigating strong spin-phonon coupling, and non-linear magneto-optical phenomena. In this work, we demonstrate an important unresolved role of spin-orbit coupling (SOC) in the ground state and excitations of this compound. Combining first principles calculations with Linear Flavor Wave Theory (LFWT), we find strong mixing and spectral overlap of different spin-orbital single-ion states. As such, the low-lying excitations form complex mixtures of local degrees of freedom most accurately viewed as hybrid spin-orbit exciton-magnon modes. Complete parameterization of the resulting low-energy model including all such degrees of freedom requires nearly half a million coupling constants. Despite this complexity, we show that such a model can be inexpensively derived using local many-body-based approaches, which yield quantitative agreement with recent experiments. The results highlight the importance of SOC even in first row transition metals, and provide essential insight into the properties of 2D magnets with unquenched orbital moments. 
\end{abstract}

\maketitle

\noindent
{\bf Introduction} 
Two-dimensional (2D) van der Waals magnets have recently been intensively investigated for fundamental developments and potential applications in heterostructure  devices \cite{burch2018magnetism,mak2019probing,li2019intrinsic,mcguire2020cleavable,kurebayashi2022magnetism}. One such material is FePS$_3$, which exhibits a large Ising anisotropy of the local moments \cite{wildes2020high,nauman2021complete}, allowing retention of (zigzag) antiferromagnetic order down to the monolayer limit \cite{lee2016ising}. This fact combined with the ease of exfoliation of few-layer samples of FePS$_3$ has facilitated recent studies of e.g.~proximity effects in magnet-semiconductor heterostructures \cite{gong2023ferromagnetism}, cavity manipulation of electromagnetic responses via strong light-matter coupling \cite{zhang2022cavity}, and giant nonlinear optical responses \cite{ni2022observation}. The material has also emerged as an ideal platform for studying strong spin-phonon coupling \cite{wang2023magnon}, as it exhibits clear formation of optical magnon-polarons \cite{liu2021direct,zhang2021coherent,vaclavkova2021magnon,sun2022magneto}, and a large modification of the lattice dynamics with magnetic state \cite{ghosh2021spin,zhou2022dynamical,ergeccen2023coherent,zong2023spin}. A similar effect leads to the possibility of topologically nontrivial phonon-magnon hybrid modes \cite{to2023giant,klogetvedt2023tunable,cui2023chirality,luo2023evidence} in FePS$_3$ and the isostructural FePSe$_3$.

\begin{figure}[t]
\includegraphics[width=0.95\linewidth]{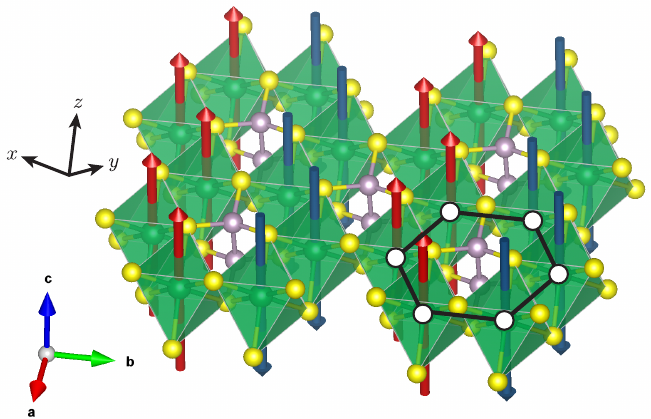}
\caption{{\bf Structure of FePS$_3$.} The (a,b,c) coordinates refer to the $C2/m$ unit cell. The (x,y,z) coordinates are the global cubic axes. Predicted ordered moment directions are indicated by red and blue arrows.}
\label{fig-structures}
\end{figure}

The bulk magnetic excitations of FePS$_3$ have also been extensively studied, via inelastic neutron scattering (INS) \cite{wildes2012magnon,lanccon2016magnetic,wildes2020evidence}, magneto-Raman \cite{mccreary2020quasi}, and THz spectroscopy \cite{wyzula2022high}. Particulary intriguing is Ref.~\onlinecite{wyzula2022high}, which proposed to observe high energy 4-magnon bound states. However, to date, the majority of such works have analyzed the response of FePS$_3$ with reference to phenomenological models with spin $S=2$, referring to the four unpaired electrons at each high-spin $d^6$ Fe$^{2+}$ site. As we discuss in this work, such models ignore the unquenched local orbital degrees of freedom, which are necessary for the strong Ising anisotropy of the local moments \cite{chandrasekharan1994magnetism,kim2021magnetic}, and play a significant role in related Fe$^{2+}$ materials \cite{song2015large,bai2021hybridized}. Experimental evidence for large orbital moments in FePS$_3$ can be seen in recent X-ray absorption spectroscopy branching ratio and linear dichroism measurements \cite{lee2023giant}. Thus, we raise a fundamental question: Is FePS$_3$ an $S = 2$ system? The definitive answer is, ``No.'' Utilizing a systematic first-principles-based approach to construct the low-energy model, we instead show that the local moments are rich mixtures of spin-orbit entangled $J_{\rm eff} = 1, 2,$ and 3 states.  The low-lying magnetic excitations alter both the orbital composition and moment orientations, thus being described as hybrid spin-orbit exciton (SOE)-magnon modes. These findings are relevant for a complete microscopic understanding of the intriguing magnetic, phononic, and optical properties of FePS$_3$.

\begin{figure}[t]
\includegraphics[width=\linewidth]{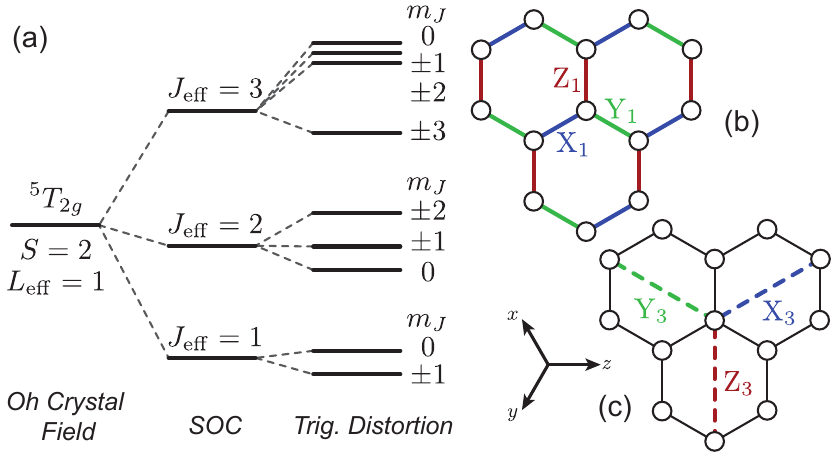}
\caption{{\bf Energy level scheme and definition of interactions.} (a) Splitting of local single-ion multiplets with spin-orbit coupling (SOC) and trigonal distortion. (b) Definition of first neighbor (X$_1$, Y$_1$, Z$_1$) bonds. (c) Definition of third neighbor (X$_3$, Y$_3$, Z$_3$) bonds. The (x,y,z) coordinates are the global cubic axes.}
\label{fig-crystalfield}
\end{figure}

\noindent
{\bf Single Site States} At the single-site level, the high-spin $d^6$ case has electronic configuration $(t_{2g})^4(e_g)^2$, which corresponds to $S = 2$, and $L_{\rm eff} = 1$. The latter orbital momentum has been largely ignored in various recent works, which treated FePS$_3$ as an $S=2$ system. Instead, when spin-orbit coupling (SOC) is considered, the low-energy single-ion states are split into $J_{\rm eff} = 1, 2,$ and 3 multiplets \cite{sanjuan1992electronic,chandrasekharan1994magnetism}, as shown in Fig.~\ref{fig-crystalfield}(a). These states are further split by the crystal field, which has approximately trigonal symmetry in FePS$_3$. The basis functions for the pure $J_{\rm eff}$ multiplets are given in [\onlinecite{sup}]. As discussed below, due to the relative weakness of SOC, intersite couplings are sufficiently strong to significantly mix and reorder these single-ion levels, such that the local moments and excitations develop a mixed $J_{\rm eff}$ character.

\noindent{\bf Electronic Hamiltonian} In order to develop a low-energy model for FePS$_3$, we use the des Cloizeaux effective Hamiltonian (dCEH) approach \cite{des1960extension,soliverez1981general} to compute interactions for each bond up to third neighbors. A description is given in the Methods section. We first consider an electronic Hamiltonian in terms of Fe $3d$-orbitals, which is a sum of, respectively, one- and two-particle terms: $\mathcal{H}_{\rm el} = \mathcal{H}_{1p}+\mathcal{H}_{2p}$. The one-particle terms include intersite hopping, intrasite crystal field, and spin-orbit coupling, $\mathcal{H}_{1p} = \mathcal{H}_{hop}+\mathcal{H}_{\rm CF} + \mathcal{H}_{\rm SO}$:
\begin{align}
    \mathcal{H}_{hop} =& \  \sum_{ij\alpha\beta\sigma}t_{ij}^{\alpha\beta} c_{i,\alpha,\sigma}^\dagger c_{j,\beta,\sigma}
    \\
    \mathcal{H}_{\rm CF} = & \ \sum_{i\alpha\beta\sigma}d_{i}^{\alpha\beta}c_{i,\alpha,\sigma}^\dagger c_{i,\beta,\sigma}
    \\
    \mathcal{H}_{\rm SO} = & \ \sum_{i\alpha\beta\sigma\sigma^\prime} \lambda_{\rm Fe} \langle \phi_i^\alpha(\sigma)|\mathbf{L}\cdot\mathbf{S}|\phi_{i}^\beta(\sigma^\prime)\rangle c_{i,\alpha,\sigma}^\dagger c_{i,\beta,\sigma^\prime}
\end{align}
where $c_{i,\alpha,\sigma}^\dagger$ creates an electron at site $i$, in orbital $\alpha$, with spin $\sigma$. To estimate $t_{ij}^{\alpha\beta}$ and $d_{i}^{\alpha\beta}$, we perform density functional theory (DFT) calculations, as described in the Methods section.

For $\mathcal{H}_{2p} = \mathcal{H}_U + \mathcal{H}_{Jnn}$, we consider both on-site and intersite terms, respectively. The on-site contributions are given by:
\begin{align}
\mathcal{H}_U = \sum_{i\alpha\beta\delta\gamma}\sum_{\sigma\sigma^\prime}U_{\alpha\beta\gamma\delta} \ c_{i,\alpha,\sigma}^\dagger c_{i,\beta,\sigma^\prime}^\dagger c_{i,\gamma,\sigma^\prime} c_{i,\delta,\sigma}
\end{align}
In the spherically symmetric approximation \cite{sugano1970multiplets}, $U_{\alpha\beta\gamma\delta}$ are parameterized by the Slater parameters $F_0, F_2, F_4$. We use $F_2 = 9.11$ eV, $F_4 = 6.56$ eV, in accordance with spectroscopic studies of high energy $d$-$d$ transitions in FePS$_3$ \cite{joy1992optical}. This leaves $F_0$ (or equivalently, $U_{t2g} = F_0 - \frac{4}{49}(F_2+F_4)$) as a free parameter in the calculation. The effects of this choice are discussed below; we ultimately conclude that $U_{t2g} = 4.2$ eV provides a good match with experiment. For the intersite terms, we consider an additional nearest neighbor Hund's coupling:
\begin{align}
    \mathcal{H}_{Jnn} = \sum_{ij\alpha\beta\sigma\sigma^\prime} J_{H,ij}^{\alpha\beta} \ c_{i,\alpha,\sigma}^\dagger c_{j,\beta,\sigma^\prime}^\dagger c_{i,\alpha,\sigma^\prime}c_{j,\beta,\sigma}
\end{align}
The physical origin of this term is the downfolding of the $p$-orbital Coulomb interactions associated with the sulfur ligands into the Fe $d$-orbital Wannier functions. The Wannier orbitals are, in reality, anti-bonding combinations of $d$- and $p$-orbitals. The usual ferromagnetic Goodenough-Kanamori superexchange \cite{kanamori1959superexchange} arises from the residual effects of the ligand Hund's coupling when $d$-orbitals of adjacent Fe atoms hybridize with $p$-orbitals of the same ligand. This term plays a primary role in establishing ferromagnetic nearest neighbor couplings in edge-sharing materials (see e.g.~\cite{autieri2022limited}). The intersite Hund's coupling coefficients $J_{H,ij}^{\alpha\beta}$ can, in principle, be estimated by constrained Random Phase Approximation (cRPA) calculations \cite{csacsiouglu2011effective,vaugier2012hubbard}. However, we have found that cRPA tends to overestimate the coefficients by an order of magnitude. Instead we take a partially empirical approach. Projecting the $p$-orbital Coulomb interactions into the Wannier function basis gives the approximation:
\begin{align}
    J_{H,ij}^{\alpha\beta} = \sum_{n,\delta,\gamma}(U^n-J_H^n) & \ \phi_{i,\alpha}^{n,\delta}\phi_{j,\beta}^{n\delta}\phi_{i,\alpha}^{n,\gamma}\phi_{j,\beta}^{n,\gamma}\nonumber \\ & \ + J_H^n|\phi_{i,\alpha}^{n,\delta}|^2 |\phi_{j,\beta}^{n,\gamma}|^2 
\end{align}
where $\phi_{i,\alpha}^{n,\delta}$ is the wavefunction coefficient for the Wannier function at Fe site $i$ and $d$-orbital $\alpha$ corresponding to the directly-bonded sulfur atom $n$, and $p$-orbital $\delta \in\{p_x,p_y,p_z\}$. $U^n$ is the Hubbard repulsion between electrons in the same $p$-orbital at site $n$, and $J_H^n$ is the Hund's coupling at site $n$. At this point, we approximate $U^n = 2 J_H^n$, and take the screened sulfur Hund's coupling to be half of the atomic value \cite{liska1975systematic}, namely $J_H^n = 0.27$ eV. We then evaluate the sums over bridging sulfur atoms, employing the wavefunction coefficients obtained from DFT. While these choices are physically reasonable, this approach is further justified because it yields intersite magnetic couplings of experimentally correct magnitude, as shown below. The full estimated intersite $J_{H,ij}^{\alpha\beta}$ tensors are given in the Supplemental Material [\onlinecite{sup}]; the largest coefficients are $\sim 2$ meV, and involve the $e_g$ orbitals, which hybridize more strongly than the $t_{2g}$ orbitals with the ligand $p$-orbitals.

\begin{figure}[t]
\includegraphics[width=0.95\linewidth]{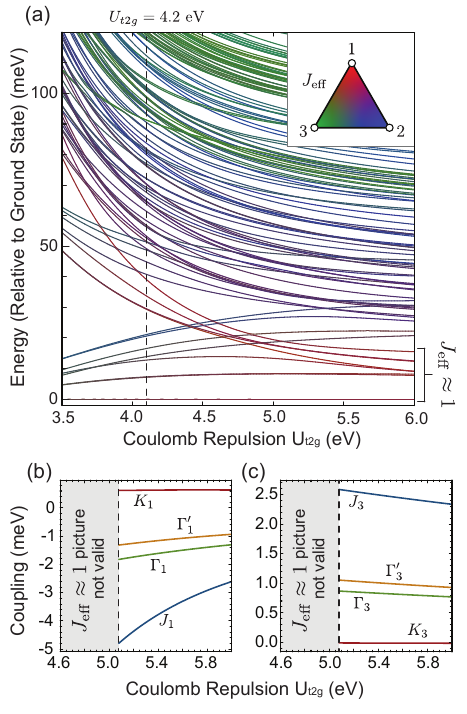}
\caption{{\bf Two-site energy levels and bilinear magnetic couplings.} (a) Evolution of lowest energy eigenvalues for two site Z$_1$-bond cluster. States are colored according to their composition in terms of $J_{\rm eff}$ states. For $U_{t2g}\lesssim 5$ eV, $J_{\rm eff} = 2$ states intrude into the low-energy space, invalidating a pure $J_{\rm eff} = 1$ model. (b,c) Evolution of bilinear couplings in $J_{\rm eff} = 1$ model for large $U_{t2g}$ for first and third neighbors, respectively. Second neighbor couplings are an order of magnitude smaller, and are not shown.}
\label{fig-udep}
\end{figure}

\noindent
{\bf Intersite Low-Energy Couplings} With the electronic Hamiltonian thus defined, we obtain the low energy model by exactly diagonalizing $\mathcal{H}_{\rm el}$ on pairs of sites up to third nearest neighbor, and projecting the lowest energy eigenstates onto pure $J_{\rm eff}$ states (see Supplemental Material [\onlinecite{sup}] for full definition). In general, the resulting low-energy model takes the form:
\begin{align}
\mathcal{H}=\sum_{i,n} A_i^n \mathcal{O}_i^n + \sum_{ij,nm}J_{ij}^{nm}\mathcal{O}_i^n \mathcal{O}_j^m
\end{align}
where $\mathcal{O}_i^n$ are local operators that act in the local basis, which may include up to $J_{\rm eff} = 1,2$, and 3 states.

In Fig.~\ref{fig-udep}(a), we first show the evolution of the state energies as a function of $U_{t2g}$ for a two-site cluster representing the Z$_1$-bond. It is instructive to first consider the large $U_{t2g}$ limit, where states of different $J$ are somewhat separated in energy. In this limit, it is possible to constrain the low-energy model to the lowest three states on each site, which are smoothly connected to pure $J_{\rm eff} = 1$ triplets (9 states for the pair of sites). In this case, the $\mathcal{O}_i^n$ operators may be chosen as the conventional spherical tensor (Stevens) operators for $S = 1$. The Hamiltonian may then be written:
\begin{align}
    \mathcal{H} = \sum_{i}& \  A (\mathbf{S}_i \cdot \hat{c}^*)^2 + \sum_{ij} \mathbf{S}_i \cdot \mathbb{J}_{ij}\cdot\mathbf{S}_j \nonumber \\ & \ + \sum_{ij,nm}B_{ij}^{nm} \ \mathcal{O}_i^{2,n} \ \mathcal{O}_j^{2,m}
\end{align}
where the bilinear couplings are parameterized by
\begin{align}
    \mathbb{J}_{ij} = \left(\begin{array}{ccc}J & \Gamma & \Gamma^\prime \\ \Gamma & J & \Gamma^\prime \\ \Gamma^\prime & \Gamma^\prime & J+K \end{array}\right)
\end{align}
for the Z-bonds, in terms of the global ($x,y,z$) coordinates defined in Fig.~\ref{fig-structures} and \ref{fig-crystalfield}(b,c). In Fig.~\ref{fig-udep}(b,c), we show the evolution of the couplings for the Z$_1$ and Z$_3$ bonds obtained by projecting the lowest 9 states onto pure $J_{\rm eff} = 1$ states. A full discussion of the biquadratic couplings $B_{ij}^{nm}$ is found in the Supplemental Material [\onlinecite{sup}]. There are several important observations: (i) We find a large single-ion anisotropy (SIA) with $A \approx -5.1$ meV, independent of $U_{t2g}$. This arises from the local trigonal distortion of the FeS$_6$ octahedra, which split the $J_{\rm eff} = 1$ levels. (ii) The intersite exchange couplings are quite anisotropic. The nearest neighbor Kitaev coupling $K_1$ is antiferromagnetic, but small compared to the other couplings. The third neighbor $K_3$ is negligible. Instead, the largest anisotropic couplings are $\Gamma_n \approx \Gamma_n^\prime$, which have the same sign as the corresponding Heisenberg couplings $J_n$. This implies a significant bond-independent Ising exchange anisotropy, with the Ising axis along the $c^*$-axis. Microscopically, the origin of this exchange anisotropy is precisely the same as the SIA: the modification of the local moments by the trigonal crystal field. Both the exchange anisotropy and SIA must be present together. (iii) There is significant anisotropic nearest neighbor biquadratic (BQ) exchange. Full details are given in the Supplemental Material [\onlinecite{sup}]; the Z-bond BQ terms are parameterized by nine symmetry-distinct constants, which we find are typically on the order of 10\% of $J_1$. It may be noted that large BQ exchange was recently implicated to explain the specific dispersion of excitations in inelastic neutron scattering experiments \cite{wildes2020evidence}, although this was in the context of an $S=2$ model. 

\begin{figure}[t]
\includegraphics[width=0.9\linewidth]{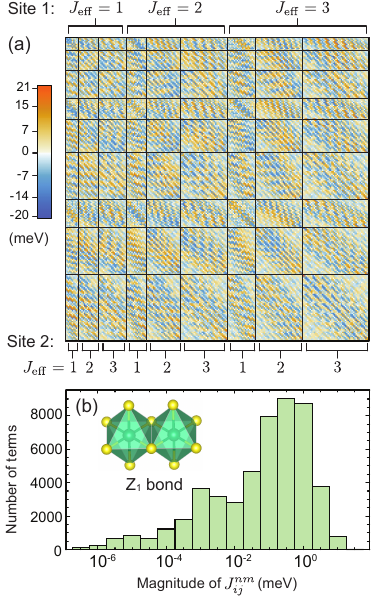}
\caption{{\bf Summary of nearest neighbor interactions for full model.} (a) Plot of coupling interaction matrix indicating location of different $J_{\rm eff}$ blocks for each site, for the Z$_1$ bond and $U_{t2g} = 4.2$ eV for full $J_{\rm eff} = 1, 2,$ and 3 model. A constant has been subtracted from the diagonal to make the matrix traceless for plotting. (b) Histogram showing distribution of the magnitude of matrix elements.}
\label{fig-histogram}
\end{figure}

While these observations for the large $U$ case may provide some intuition into the intersite couplings, the $J_{\rm eff} = 1$ picture breaks down for $U_{t2g}\lesssim 5$ eV, as the higher lying $J_{\rm eff} = 2$ states descend into the low-energy window. This is depicted in Fig.~\ref{fig-udep}(a). As we show below, the physically applicable value is $U_{t2g} \approx 4.2 < 5$ eV. Thus, it is not possible to map the low-energy space onto a single $J$ multiplet,  so that all 15 local $J_{\rm eff}=1,2,$ and 3 states need to be considered explicitly in the low-energy model. 
Complete specification of the Hamiltonian therefore requires $15^2 = 225$ local operators, such that each bond interaction is defined by $(15^2-1)^2=50,176$ coupling constants $J_{ij}^{nm}$. Fig.~\ref{fig-histogram}(a) and (b) show the distribution and magnitudes of such couplings for the Z$_1$ bond obtained by the dCEH method for $U_{t2g} = 4.2$ eV; the fact that there is no clear separation of magnitudes implies the need to retain all couplings. This situation highlights the utility of many-body approaches, such as the above-described dCEH approach, for the estimation of low-energy Hamiltonians (see e.g.~\cite{calzado2002analysis,calzado2002analysis2,pourovskii2016two,riedl2019ab,pourovskii2021ferro} for a survey of approaches). The many-body nature of the local spin-orbital states and large number of couplings would present a significant difficulty for more traditional pure-DFT approaches in which couplings are estimated from energy differences between suitably constrained Kohn-Sham (single determinant) wavefunctions \cite{pi2014calculation,mosca2022modeling,szilva2023quantitative}. In the above-described dCEH approach, all couplings are instead obtained simultaneously for a given bond from a single diagonalization of the two-site electronic Hamiltonian, which requires at most a few minutes on a single workstation. In lieu of printing these couplings, we provide the corresponding bond-matrices in matrix market \cite{boisvert1996matrix} (.mtx) format in the Supplemental Material [\onlinecite{sup}] to facilitate future studies.

\begin{figure*}[t]
\includegraphics[width=\linewidth]{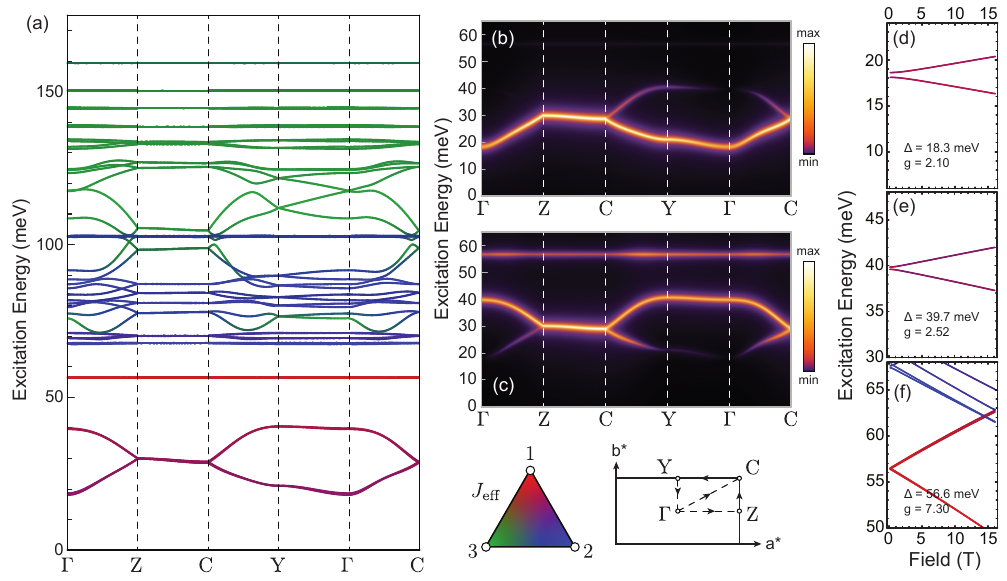}
\caption{{\bf Computed Excitations from Linear Flavor Wave Theory.} (a) Full magnon/spin-orbit exciton band structure, with bands colored according to their $J_{\rm eff}$ composition. (b) Dynamical magnetic structure factor $D_{\rm mag}(q)$ for the lowest energy bands. (c) Dynamical electrical polarization structure factor $D_{\rm el}(q)$ for the lowest energy bands (see text for definition). (d-f) Evolution of the lowest excitations with magnetic field oriented along the $c^*$-axis, with zero-field energy $\Delta$ and slope $g$ indicated.  }
\label{fig-bands}
\end{figure*}

\noindent
{\bf Model Properties} From here, we consider only the full $J_{\rm eff} = 1,2,3$ model with $U_{t2g} = 4.2$ eV. 
In order to analyze the ground state and excitations, we employ a numerical Linear Flavor Wave Theory (LFWT) approach \cite{joshi1999elementary,lauchli2006quadrupolar,toth2010three,luo2016spin,kim2017linear,koyama2023flavor}, described in detail in the Methods section. We find that the full model reproduces {\it all} aspects of the experimental response of FePS$_3$. 

In agreement with experiment, the mean field model exhibits antiferromagnetic zigzag order of magnetic dipoles, which can be understood as arising from primarily ferromagnetic nearest neighbor and antiferromagnetic third neighbor couplings. The moment size is $|\langle\psi_i^0|\mathbf{L}-2\mathbf{S}|\psi_i^0\rangle| = 5.0 \mu_B$, which is enhanced compared to the spin-only value of $4 \mu_B$ for $S=2$. A similar enhancement has been reported both from neutron \cite{kurosawa1983neutron,coak2021emergent} and susceptibility \cite{joy1992magnetism} analysis.

For the zigzag wavevector along the $b$-axis, the moments are oriented in the $ac^*$-plane, making an angle of 11$^\circ$ with the $c^*$ axis, as shown in Fig.~\ref{fig-structures}. Previous neutron scattering data was analyzed in terms of the moments oriented precisely along the $c^*$ axis, although this is not required by symmetry. Future experiments could refine the moment direction. The composition of the MF ground state is 75\% $J_{\rm eff}=1$, 23\% $J_{\rm eff}=2$, and 2\% $J_{\rm eff}=3$, demonstrating significant mixture of different spin-orbital components. This mixing is driven both by the on-site crystal field and the (mean-field) effects of intersite coupling.  

In principle, there are two classes of excitations of the magnetically ordered phase: magnon-like modes within the lowest $J_{\rm eff}$ manifold, and spin-orbit excitons (SOE) between different $J_{\rm eff}$ values. In FePS$_3$, we find that these excitations are strongly mixed. The predicted dispersion of the hybrid SOE/magnon excitations is shown in Fig.~\ref{fig-bands}(a), with the color indicating the $J$-composition of the mode. In Fig.~\ref{fig-bands}(b), we show the predicted low-energy dynamical magnetic structure factor, defined by $D_{\rm mag}(q) = \sum_\mu \int e^{i\omega t} [L_{-q}^\mu (t) - 2S_{-q}^\mu (t) ][ L_q^\mu(0) - 2S_q^\mu(0)] dt$. The intensity pattern matches well with the inelastic neutron scattering (INS) experiments \cite{wildes2012magnon,lanccon2016magnetic,wildes2020evidence}. 

In accordance with INS experiments, the lowest branch forms two dispersive bands. We predict the gap at the $\Gamma$-point to be $\sim 18$ meV, which is somewhat larger than the experimental value of 15 meV. This discrepancy is likely due to an overestimation of the trigonal crystal field splitting at the DFT level. For example, a recent x-ray absorption study \cite{lee2023giant} was well-modelled with an absolute splitting of the $t_{2g}$ orbitals on the order of $\sim 10$ meV, while we find a value of $\sim 44$ meV from the Wannier-interpolated crystal field terms. Apart from this discrepancy, the predicted dispersions of the excitations below 50 meV are in excellent agreement with INS experiments. Due to their differing spin-orbital composition, population of the low-energy modes alters the $J$-composition of the local moments. 
For example, at $q=0$, the eigenvector for the lower band corresponds to a composition of 53\% $J_{\rm eff}=1$, 38\% $J_{\rm eff}=2$, and 9\% $J_{\rm eff}=3$. Similarly, the upper band has a composition of 65\% $J_{\rm eff}=1$, 28\% $J_{\rm eff}=2$, and 7\% $J_{\rm eff}=3$.
In both cases, relative weight is shifted from $J_{\rm eff} = 1$ to $J_{\rm eff} = 2$, highlighting the mixed SOE/magnon character of these modes. Another particular feature of the bands is the very weak dispersion along the $\Gamma \to$ Y and Z $\to$ C paths, which implies the excitations primarily propagate perpendicular to the ordering wavevector, i.e.~along the ferromagnetic zigzag chains. This feature was discussed in Ref.~\onlinecite{wildes2020evidence}, and is naturally reproduced here.

Above the lowest branch, there is a flat band at $\sim 57$ meV. This band essentially represents the quadrupolar $\Delta m_J = \pm 2$ modes of the lowest $J_{\rm eff} = 1$ single-ion level. These modes correspond to a composition of 78\% $J_{\rm eff}=1$, 13\% $J_{\rm eff}=2$, and 9\% $J_{\rm eff}=3$, which is similar to the composition of the ground-state moments. Beyond the LFWT approximation, we would expect this mode to hybridize with the 2-magnon continuum, which may lead to broadening. Finally, at higher energies there are a number of modes between 70 and 170 meV, with mostly $J_{\rm eff} = 2$ and 3 SOE character. These likely explain the broad absorption bands observed in the same region of energies via infrared and Raman measurements \cite{jouanne1988fourier,sanjuan1992electronic}.

We next consider the THz/infrared studies reported in Ref.~\cite{wyzula2022high}. These studies observed the main modes at 15.1, 39.6, and 57.5 meV, which correspond very well with the predicted energies of 18.3, 39.7, and 56.6 meV from LFWT (at $q = 0$). In Fig.~\ref{fig-bands}(d-f), we show the evolution of these modes with magnetic field applied along the out-of-plane $c^*$ direction. The modes at 18.3 and 39.7 meV each consist of a pair of excitations that are symmetrically split under applied field. These represent magnon-like excitations ($\Delta m_J \sim \pm 1$), which are essentially confined to either a spin-up or spin-down sublattice. Based on the predicted slopes, we evaluate the effective $g$-values for the excitations to be 2.10 and 2.52, respectively. These values are in good agreement with the experimental value of 2.15 for both modes. The quadrupolar ($\Delta m_J \sim \pm 2$) mode at 56.6 meV instead has a much larger slope, which we evaluate as $g \approx 7.3$. Experimentally, it was found to be 9.2. While, naively, one might expect the energies of the $\Delta m_J \sim \pm 2$ excitations to have twice the slope of the $\Delta m_J \sim \pm 1$ excitations with respect to field, it should be emphasized that the spin-orbital composition of the excitations is quite different. The lower energy modes have significantly higher $J_{\rm eff} = 2$ character, which reduces the effective $g$-value.

Finally, it is worth considering why the 39.7 meV mode has appreciable intensity in the THz/infrared spectra? As shown in Fig.~\ref{fig-bands}(b), this band has vanishing weight in the dynamical magnetic structure factor near $q=0$. This is because this mode is odd under inversion, while the magnetic dipole operator is even. In principle, the finite intensity may therefore reflect a lowering of symmetry in the bulk, or a surface effect in thin samples. Here we present another possibility that arises from the orbital component of the excitations: in a sufficiently low-symmetry crystal field, these modes are electric dipole-active. In order to demonstrate this, we computed the electrical polarization operators in the dCEH basis from the matrix elements of $\mathbf{P} = -e\mathbf{r}$ between $d$-orbital Wannier functions on the same Fe site. In Fig.~\ref{fig-bands}(c), we show the predicted electric dipole structure factor, defined by $D_{\rm el}(q) = \sum_\mu \int e^{i\omega t} P_{-q}^\mu (t) P_q^\mu(0) dt$. For $q=0$, this is proportional to the electric dipole absorption intensity, which contributes to THz/infrared spectra, but not INS. Since $\mathbf{P}$ is odd under inversion, this provides a route to observe the inversion-odd ``exchange modes'' without lowering of symmetry, provided such excitations significantly alter the orbital composition. In the present case of FePS$_3$, where the magnons and SOEs are strongly mixed, this condition is satisfied. As shown in Fig.~\ref{fig-bands}(c), both the 39.7 and 56.6 meV excitations are electric dipole active.

{\bf Conclusions} In this work, we have investigated the low-energy model, ground state, and excitations of FePS$_3$, which has been of growing interest in the context of 2D magnetism. Contrary to typical expectations for $3d$ transition metal compounds, SOC is very important to accurately model the low-lying excitations. In fact, the weakness of SOC compared to crystal field splitting and intersite exchange leads to spectral overlap and mixing of magnons and spin-orbit excitons to form hybrid excitations, the properties of which depend on their specific spin-orbital composition. We have shown that a comprehensive model including all local $J_{\rm eff}$ states, obtained from first principles calculations employing the dCEH approach, reproduces all essential experimental details with remarkable quantititive agreement. While these findings do not invalidate previous phenomenological models (of e.g. spin-phonon coupling in FePS$_3$) in terms of an $S=2$ system, a complete microscopic picture of the intriguing optical, magnetic, and phononic properties should account for the large orbital moment. For example, due to the relative weakness of SOC, the specific spin-orbital composition of the local moments and excitations can be strongly perturbed by structural distortions, potentially enhancing spin-lattice coupling and sensitivity to symmetry breaking at the surface. Effects such as the giant surface optical second harmonic generation \cite{ni2022observation}, and strong spin-phonon coupling \cite{liu2021direct,zhang2021coherent,vaclavkova2021magnon,sun2022magneto} may therefore be rooted in SOC. Overall, the results highlight that complex spin-orbital phenomena may be found even in first row transition metal compounds, and that we have the tools to address them.

{\bf Methods} {\it Density Functional Theory} Ab-initio DFT calculations were performed starting from the  experimental $C2/m$ structure of FePS$_3$ \cite{ouvrard1985structural} in order to parameterize the single-particle parts of the electronic Hamiltonian. These calculations employed the package FPLO \cite{koepernik1999full,opahle1999full}, and were performed at the GGA (PBE) level \cite{perdew1996generalized} using a $12\times 12\times 12$ $k$-grid for self-consistent calculations. Hopping integrals were obtained by projecting the resulting electronic bands onto Fe $d$-orbitals to construct Wannier functions \cite{koepernik2023symmetry}. We also repeated this procedure for fully relativistic calculations in order to check the effects of SOC on the single-particle Wannier Hamiltonian. From the comparison of the two calculations, we find that SOC is well captured by an on-site $\lambda \mathbf{L}\cdot\mathbf{S}$ term with $\lambda = 53$ meV, the atomic value for Fe \cite{montalti2006handbook}, which was subsequently included in the derivation of the low-energy Hamiltonian. Imaginary intersite hoppings were negligible in the fully relativistic calculation, and were subsequently omitted.

{\it des Cloizeaux effective Hamiltonians} Calculations of the low-energy Hamiltonians were performed using the dCEH approach  \cite{des1960extension,soliverez1981general}. This is a standard method for extracting generic effective Hamiltonians, which act within an idealized low-energy Hilbert space (e.g.~spanned by local spin/orbital degrees of freedom), from numerical eigenstates of a larger Hilbert space (e.g.~a full electronic Fock space). In our implementation, the electronic Hamiltonian was first diagonalized on a finite number of Fe sites to obtain the low-energy eigenstates $|\phi_n\rangle$ and energies $E_n$ in terms of the electronic degrees of freedom. The low-energy Hamiltonian is then formally $\mathcal{H}_{\rm low} = \sum_n E_n |\phi_n\rangle \langle \phi_n|$. However, in order to interpret this expression, a mapping is required between the electronic states and the idealized low energy spin-orbital states, $|\psi_n\rangle$. As described in the Results section, this idealized basis may be chosen as pure $J_{\rm eff} = 1$ states at each site in the case of large $U$, but must include all $J_{\rm eff} = 1,2$, and 3 states for realistic values of $U$ due to energetic overlap of different angular momenta. $\mathcal{H}_{\rm low}$ was therefore rotated into the idealized low-energy space, via $ \mathcal{H}_{\rm eff} = \mathbf{S}^{-1/2} \mathbb{P} \mathcal{H}_{\rm low} \mathbb{P}\mathbf{S}^{-1/2}$, where  $\mathbb{P} = \sum_n |\psi_n\rangle\langle \psi_n|$ is the projection operator onto the low-energy space, and $\mathbf{S}$ is the overlap matrix of the projected low-energy states, with matrix elements $ \left[\mathbf{S}\right]_{nm} = \langle \phi_n| \mathbb{P}|\phi_m\rangle$. It may be noted that $U = \mathbb{P}\mathbf{S}^{-1/2}$ is a unitary operator (with $U^\dagger = \mathbf{S}^{-1/2}\mathbb{P}$), so this transformation preserves the local spectrum of $\mathcal{H}_{\rm low}$. For evaluation of expectation values, such as the magnetization or dipole matrix elements, the operators within the electronic Hilbert space were transformed accordingly, i.e. $\mathbf{L}-2\mathbf{S} \to U^\dagger(\mathbf{L}-2\mathbf{S})U$. This ensures the correct representation of these operators within the low-energy effective model.

{\it Linear Flavor Wave Theory} Calculations of the ground state and excitations of FePS$_3$ were performed within the LFWT approach. The mean-field model $\mathcal{H}_{\rm MF} = \sum_{i,n}\left(A_i^n + \sum_{jm} J_{ij}^{nm}\langle \mathcal{O}_j^m\rangle\right) \mathcal{O}_i^n$ was first solved self-consistently. From this, the local eigenstates $|\psi_i^\alpha\rangle$ were then employed to construct the LFWT Hamiltonian in terms of bosonic operators $b_{i,\alpha}^\dagger \equiv |\psi_i^\alpha\rangle\langle\psi_i^0|$
where $|\psi_i^0\rangle$ corresponds to the local ground state of the MF model at site $i$. Then, we may write:
\begin{align}
    \mathcal{H}_{\rm LFWT} =& \  \sum_i \varepsilon_i^\alpha b_{i,\alpha}^\dagger b_{i,\alpha} + \sum_{ij} t_{ij}^{\alpha\beta} b_{i,\alpha}^\dagger b_{j,\beta} \nonumber \\ & \  + \sum_{ij}\Delta_{ij}^{\alpha\beta} b_{i,\alpha}^\dagger b_{j,\beta}^\dagger + H.c.
\end{align}
where:
\begin{align}
    \varepsilon_i^\alpha =& \  \langle \psi_i^\alpha| \mathcal{H}_{\rm MF} | \psi_i^\alpha\rangle
    \\
    t_{ij}^{\alpha\beta} = & \ J_{ij}^{nm} \langle \psi_i^\alpha| \mathcal{O}_i^n | \psi_i^0\rangle\langle \psi_j^0| \mathcal{O}_j^m | \psi_j^\beta\rangle
    \\
    \Delta_{ij}^{nm} = & \ J_{ij}^{nm} \langle \psi_i^\alpha| \mathcal{O}_i^n | \psi_i^0\rangle\langle \psi_j^\beta| \mathcal{O}_j^m | \psi_j^0\rangle
\end{align}
The linearized model was then diagonalized utilizing the standard Cholesky decomposition approach \cite{colpa1978diagonalization,toth2015linear}. Following a similar approach, operators corresponding to the spin, orbital momentum, and electric dipole operators were constructed in terms of the bosonic operators $b_{i,\alpha}^\dagger$, and the $q$-dependent excitation intensities computed from the resulting eigenvectors. 

{\bf Data Availability} All data generated or analysed during this study are included in this published article and its supplementary information files.

\begin{acknowledgments}
{\bf Acknowledgements} The authors would like to thank M. Ozerov for many insightful discussions on FePS$_3$ and other materials. This research was funded by the Center for Functional Materials at WFU through a pilot grant, and Oak Ridge Associated Universities (ORAU) through the Ralph E. Powe Junior Faculty Enhancement Award to S.M.W. Computations were performed using the Wake Forest University (WFU) High Performance Computing
Facility \cite{WakeHPC}, a centrally managed computational resource available to WFU researchers including
faculty, staff, students, and collaborators.
\end{acknowledgments}

{\bf Author Contributions} S.M.W. designed the research. R.D. and S.G. performed the numerical computations. S.M.W. wrote the manuscript with input from all authors. 

{\bf Competing Interests} The authors declare no competing interests.

\bibliographystyle{naturemag}
\bibliography{feps3}
    
    \clearpage
    
    \section{Supplemental Material}
    \subsection{Basis functions for J States}
    In order to obtain the effective Hamiltonian describing the $J_{\rm eff}=1,2$, and 3 states, it is necessary to project the low-energy eigenstates of a given two-site cluster onto ideal reference states, according to the standard dCEH procedure. These states can be easily derived, but are reproduced here in full detail for future convenience. The states labelled according to $|J_{\rm eff}, m_J\rangle$ can be expanded in terms of $|S,m_S;L_{\rm eff}, m_L\rangle$ states. In particular, the $J_{\rm eff} = 3$ states are:
    \begin{align}
    |3,+3\rangle = & \ |2,+2;1,+1\rangle
    \\
    |3,+2\rangle = & \  \sqrt{\frac{1}{3}}|2,+2;1,0\rangle+\sqrt{\frac{2}{3}}|2,+1;1,+1\rangle
    \\
    |3,+1\rangle = & \ \sqrt{\frac{1}{15}}|2,+2;1,-1\rangle + \sqrt{\frac{8}{15}}|2,+1;1,0\rangle \nonumber \\ & \ \ \ \ + \sqrt{\frac{6}{15}}|2,0;1,+1\rangle
    \\
    |3,0\rangle = & \ \sqrt{\frac{1}{5}}|2,+1;1,-1\rangle + \sqrt{\frac{3}{5}}|2,0;1,0\rangle \nonumber \\ & \ \ \ \ + \sqrt{\frac{1}{5}}|2,-1;1,+1\rangle
    \\
    |3,-1\rangle = & \ \sqrt{\frac{1}{15}}|2,-2;1,+1\rangle + \sqrt{\frac{8}{15}}|2,-1;1,0\rangle \nonumber \\ & \ \ \ \ + \sqrt{\frac{6}{15}}|2,0;1,-1\rangle
    \\
    |3,-2\rangle = & \ \sqrt{\frac{1}{3}}|2,-2;1,0\rangle+\sqrt{\frac{2}{3}}|2,-1;1,-1\rangle
    \\
    |3,-3\rangle = & \ |2,-2;1, -1\rangle
    \end{align}
    The $J_{\rm eff} = 2$ states are:
    \begin{align}
    |2,+2\rangle = & \  \sqrt{\frac{2}{3}}|2,+2,1,0\rangle-\sqrt{\frac{1}{3}}|2,+1,1,+1\rangle
    \\
    |2,+1\rangle = & \ \sqrt{\frac{1}{3}}|2,+2,1,-1\rangle + \sqrt{\frac{1}{6}}|2,+1,1,0\rangle \nonumber \\ & \ \ \ \ - \sqrt{\frac{1}{2}}|2,0,1,+1\rangle
    \\
    |2,0\rangle = & \ \sqrt{\frac{1}{2}}|2,+1,1,-1\rangle  - \sqrt{\frac{1}{2}}|2,-1,1,+1\rangle
    \\
    |2,-1\rangle = & \ -\sqrt{\frac{1}{3}}|2,-2,1,+1\rangle - \sqrt{\frac{1}{6}}|2,-1,1,0\rangle \nonumber \\ & \ \ \ \ + \sqrt{\frac{1}{2}}|2,0,1,-1\rangle
    \\
    |2,-2\rangle = & \ -\sqrt{\frac{2}{3}}|2,-2,1,0\rangle+\sqrt{\frac{1}{3}}|2,-1,1,-1\rangle
    \end{align}
    The $J_{\rm eff}=1$ states are:
    \begin{align}
    |1,+1\rangle = & \ \sqrt{\frac{3}{5}}|2,+2,1,-1\rangle - \sqrt{\frac{3}{10}}|2,+1,1,0\rangle \nonumber \\ & \ \ \ \ + \sqrt{\frac{1}{10}}|2,0,1,+1\rangle
    \\
    |1,0\rangle = & \ \sqrt{\frac{3}{10}}|2,+1,1,-1\rangle - \sqrt{\frac{2}{5}}|2,0,1,0\rangle \nonumber \\ & \ \ \ \ + \sqrt{\frac{3}{10}}|2,-1,1,+1\rangle
    \\
    |1,-1\rangle = & \ \sqrt{\frac{3}{5}}|2,-2,1,+1\rangle -\sqrt{\frac{3}{10}}|2,-1,1,0\rangle \nonumber \\ & \ \ \ \ + \sqrt{\frac{1}{10}}|2,0,1,-1\rangle
    \end{align}
    Each of the $|S,m_S;L_{\rm eff}, m_L\rangle$ states can then be written as a linear combination of Slater determinants. For this purpose, we define the following orbitals:
    \begin{align}
    e_a = d_{z^2} \ \ \ , \ \ \ e_b = d_{x^2-y^2} \ \ \ , \ \ \ t_0 = d_{xy}\\
    t_+ = -\frac{1}{\sqrt{2}}(d_{yz}+id_{xz}) \ \ \ , \ \ \ t_- = \frac{1}{\sqrt{2}}(d_{yz}-id_{xz})
    \end{align}
    Then we have:
    \begin{align}
    |2,+2,1,+1\rangle = & \ |e_{a\uparrow}e_{b\uparrow}t_{+\uparrow}t_{+\downarrow}t_{0\uparrow}t_{-\uparrow}\rangle
    \\
    |2,+2,1,0\rangle = & \ -|e_{a\uparrow}e_{b\uparrow}t_{+\uparrow}t_{0\uparrow}t_{0\downarrow}t_{-\uparrow}\rangle
    \\
    |2,+2,1,-1\rangle = & \ |e_{a\uparrow}e_{b\uparrow}t_{+\uparrow}t_{0\uparrow}t_{-\uparrow}t_{-\downarrow}\rangle
    \end{align}
    \begin{align}
    |2,+1,1,+1\rangle = & \ \frac{1}{2}\left( |e_{a\downarrow}e_{b\uparrow}t_{+\uparrow}t_{+\downarrow}t_{0\uparrow}t_{-\uparrow}\rangle\right. \nonumber \\
    & \ \left. +|e_{a\uparrow}e_{b\downarrow}t_{+\uparrow}t_{+\downarrow}t_{0\uparrow}t_{-\uparrow}\rangle\right. \nonumber \\
    & \ \left. + |e_{a\uparrow}e_{b\uparrow}t_{+\uparrow}t_{+\downarrow}t_{0\downarrow}t_{-\uparrow}\rangle\right. \nonumber \\
    & \ \left. +|e_{a\uparrow}e_{b\uparrow}t_{+\uparrow}t_{+\downarrow}t_{0\uparrow}t_{-\downarrow}\rangle
    \right)
    \\
    |2,+1,1,0\rangle = & \ -\frac{1}{2}\left( |e_{a\downarrow}e_{b\uparrow}t_{+\uparrow}t_{0\uparrow}t_{0\downarrow}t_{-\uparrow}\rangle\right. \nonumber \\
    & \ \left. + |e_{a\uparrow}e_{b\downarrow}t_{+\uparrow}t_{0\uparrow}t_{0\downarrow}t_{-\uparrow}\rangle
    \right. \nonumber \\
    & \ \left. +|e_{a\uparrow}e_{b\uparrow}t_{+\downarrow}t_{0\uparrow}t_{0\downarrow}t_{-\uparrow}\rangle\right. \nonumber \\
    & \ \left. +|e_{a\uparrow}e_{b\uparrow}t_{+\uparrow}t_{0\uparrow}t_{0\downarrow}t_{-\downarrow}\rangle
    \right)
    \\
    |2,+1,1,-1\rangle = & \ \frac{1}{2}\left( |e_{a\downarrow}e_{b\uparrow}t_{+\uparrow}t_{0\uparrow}t_{-\uparrow}t_{-\downarrow}\rangle\right. \nonumber \\
    & \ \left. +|e_{a\uparrow}e_{b\downarrow}t_{+\uparrow}t_{0\uparrow}t_{-\uparrow}t_{-\downarrow}\rangle
    \right. \nonumber \\
    & \ \left. +|e_{a\uparrow}e_{b\uparrow}t_{+\downarrow}t_{0\uparrow}t_{-\uparrow}t_{-\downarrow}\rangle\right. \nonumber \\
    & \ \left. +|e_{a\uparrow}e_{b\uparrow}t_{+\uparrow}t_{0\downarrow}t_{-\uparrow}t_{-\downarrow}\rangle
    \right)
    \end{align}
    \begin{align}
    |2,0,1,+1\rangle = & \ \frac{1}{\sqrt{6}}\left( |e_{a\downarrow}e_{b\downarrow}t_{+\uparrow}t_{+\downarrow}t_{0\uparrow}t_{-\uparrow}\rangle\right.\nonumber \\
    & \ \left. +|e_{a\downarrow}e_{b\uparrow}t_{+\uparrow}t_{+\downarrow}t_{0\downarrow}t_{-\uparrow}\rangle\right.\nonumber \\
    & \ \left. +|e_{a\downarrow}e_{b\uparrow}t_{+\uparrow}t_{+\downarrow}t_{0\uparrow}t_{-\uparrow}\rangle\right. \nonumber \\
    & \ \left. +|e_{a\uparrow}e_{b\downarrow}t_{+\uparrow}t_{+\downarrow}t_{0\uparrow}t_{-\downarrow}\rangle\right.\nonumber \\
    & \ \left. + |e_{a\uparrow}e_{b\downarrow}t_{+\uparrow}t_{+\downarrow}t_{0\downarrow}t_{-\uparrow}\rangle\right. \nonumber \\
    & \ \left. +|e_{a\uparrow}e_{b\uparrow}t_{+\uparrow}t_{+\downarrow}t_{0\downarrow}t_{-\downarrow}\rangle
    \right)
    \\
    |2,0,1,0\rangle = & \ - \frac{1}{\sqrt{6}}\left( |e_{a\downarrow}e_{b\downarrow}t_{+\uparrow}t_{0\uparrow}t_{0\downarrow}t_{-\uparrow}\rangle\right. \nonumber \\
    & \ \left. +|e_{a\downarrow}e_{b\uparrow}t_{+\downarrow}t_{0\uparrow}t_{0\downarrow}t_{-\uparrow}\rangle\right. \nonumber \\
    & \ \left. +|e_{a\downarrow}e_{b\uparrow}t_{+\uparrow}t_{0\uparrow}t_{0\downarrow}t_{-\downarrow}\rangle\right. \nonumber \\
    & \ \left. +|e_{a\uparrow}e_{b\downarrow}t_{+\downarrow}t_{0\uparrow}t_{0\downarrow}t_{-\uparrow}\rangle\right. \nonumber \\
    & \ \left. + |e_{a\uparrow}e_{b\downarrow}t_{+\uparrow}t_{0\uparrow}t_{0\downarrow}t_{-\downarrow}\rangle\right. \nonumber \\
    & \ \left. + |e_{a\uparrow}e_{b\uparrow}t_{+\downarrow}t_{0\uparrow}t_{0\downarrow}t_{-\downarrow}\rangle
    \right)
    \\
    |2,0,1,-1\rangle = & \ \frac{1}{\sqrt{6}}\left( |e_{a\downarrow}e_{b\downarrow}t_{+\uparrow}t_{0\uparrow}t_{-\uparrow}t_{-\downarrow}\rangle\right. \nonumber \\
    & \ \left. + |e_{a\downarrow}e_{b\uparrow}t_{+\downarrow}t_{0\uparrow}t_{-\uparrow}t_{-\downarrow}\rangle\right. \nonumber \\
    & \ \left. + |e_{a\downarrow}e_{b\uparrow}t_{+\uparrow}t_{0\downarrow}t_{-\uparrow}t_{-\downarrow}\rangle\right. \nonumber \\
    & \ \left. + |e_{a\uparrow}e_{b\downarrow}t_{+\downarrow}t_{0\uparrow}t_{-\uparrow}t_{-\downarrow}\rangle\right. \nonumber \\
    & \ \left. + |e_{a\uparrow}e_{b\downarrow}t_{+\uparrow}t_{0\downarrow}t_{-\uparrow}t_{-\downarrow}\rangle\right. \nonumber \\
    & \ \left. + |e_{a\uparrow}e_{b\uparrow}t_{+\downarrow}t_{0\downarrow}t_{-\uparrow}t_{-\downarrow}\rangle
    \right)
    \end{align}
    \begin{align}
    |2,-1,1,+1\rangle = & \ \frac{1}{2}\left( |e_{a\uparrow}e_{b\downarrow}t_{+\uparrow}t_{+\downarrow}t_{0\downarrow}t_{-\downarrow}\rangle\right. \nonumber \\
    & \ \left. + |e_{a\downarrow}e_{b\uparrow}t_{+\uparrow}t_{+\downarrow}t_{0\downarrow}t_{-\downarrow}\rangle\right. \nonumber \\
    & \ \left. + |e_{a\downarrow}e_{b\downarrow}t_{+\uparrow}t_{+\downarrow}t_{0\uparrow}t_{-\downarrow}\rangle\right. \nonumber \\
    & \ \left. +|e_{a\downarrow}e_{b\downarrow}t_{+\uparrow}t_{+\downarrow}t_{0\downarrow}t_{-\uparrow}\rangle
    \right)
    \\
    |2,-1,1,0\rangle = & \ -\frac{1}{2}\left( |e_{a\uparrow}e_{b\downarrow}t_{+\downarrow}t_{0\uparrow}t_{0\downarrow}t_{-\downarrow}\rangle\right. \nonumber \\
    & \ \left. +|e_{a\downarrow}e_{b\uparrow}t_{+\downarrow}t_{0\uparrow}t_{0\downarrow}t_{-\downarrow}\rangle\right. \nonumber \\
    & \ \left. +|e_{a\downarrow}e_{b\downarrow}t_{+\uparrow}t_{0\uparrow}t_{0\downarrow}t_{-\downarrow}\rangle\right. \nonumber \\
    & \ \left. +|e_{a\downarrow}e_{b\downarrow}t_{+\downarrow}t_{0\uparrow}t_{0\downarrow}t_{-\uparrow}\rangle
    \right)
    \\
    |2,-1,1,-1\rangle = & \ \frac{1}{2}\left( |e_{a\uparrow}e_{b\downarrow}t_{+\downarrow}t_{0\downarrow}t_{-\uparrow}t_{-\downarrow}\rangle\right. \nonumber \\
    & \ \left. +|e_{a\downarrow}e_{b\uparrow}t_{+\downarrow}t_{0\downarrow}t_{-\uparrow}t_{-\downarrow}\rangle\right. \nonumber \\
    & \ \left. +|e_{a\downarrow}e_{b\downarrow}t_{+\uparrow}t_{0\downarrow}t_{-\uparrow}t_{-\downarrow}\rangle\right. \nonumber \\
    & \ \left. +|e_{a\downarrow}e_{b\downarrow}t_{+\downarrow}t_{0\uparrow}t_{-\uparrow}t_{-\downarrow}\rangle
    \right)
    \\
    |2,-2,1,+1\rangle = & \ |e_{a\downarrow}e_{b\downarrow}t_{+\uparrow}t_{+\downarrow}t_{0\downarrow}t_{-\downarrow}\rangle
    \\
    |2,-2,1,0\rangle = & \ -|e_{a\downarrow}e_{b\downarrow}t_{+\downarrow}t_{0\uparrow}t_{0\downarrow}t_{-\downarrow}\rangle
    \\
    |2,-2,1,-1\rangle = & \ |e_{a\downarrow}e_{b\downarrow}t_{+\downarrow}t_{0\downarrow}t_{-\uparrow}t_{-\downarrow}\rangle
    \end{align}
    
    \subsection{Intersite Hund's Coupling Matrices}
    As discussed in the main text, we include the effects of intersite Hund's coupling for nearest neighbor bonds, in order to treat the ferromagnetic Goodenough-Kanamori exchange. The resulting electronic Hamiltonian takes the form:
    \begin{align}
    \mathcal{H}_{Jnn} = \sum_{ij\alpha\beta\sigma\sigma^\prime} J_{H,ij}^{\alpha\beta} \ c_{i,\alpha,\sigma}^\dagger c_{j,\beta,\sigma^\prime}^\dagger c_{i,\alpha,\sigma^\prime}c_{j,\beta,\sigma}
    \end{align}
    where the coefficients $J_{H,ij}^{\alpha\beta}$ were approximated as:
    \begin{align}
    J_{H,ij}^{\alpha\beta} = \sum_{n,\delta,\gamma}(U^n-J_H^n) & \ \phi_{i,\alpha}^{n,\delta}\phi_{j,\beta}^{n\delta}\phi_{i,\alpha}^{n,\gamma}\phi_{j,\beta}^{n,\gamma}\nonumber \\ & \ + J_H^n|\phi_{i,\alpha}^{n,\delta}|^2 |\phi_{j,\beta}^{n,\gamma}|^2 
    \end{align}
    with $U^n = 2 J_H^n$, and $J_H^n = 0.27$ eV. The resulting tensors are thus summarized as (in meV):
    
    X$_1$ bond:
    \begin{align}
    J_{H,ij}^{\alpha\beta} = \left(\begin{array}{r|ccccc}
    &d_{xy}&d_{xz}&d_{yz}&d_{z^2}&d_{x^2-y^2}\\
    \hline
    d_{xy}& 0. & 0.03 & 0.08 & 0.2 & 0.02\\
    d_{xz}& 0.03 & 0. & 0.09 & 0.07 & 0.16\\
    d_{yz}& 0.08 & 0.09 & 0.44 & 1.2 & 0.8\\
    d_{z^2}& 0.2 & 0.07 & 1.2 & 0.7 & 1.32\\
    d_{x^2-y^2} & 0.02 & 0.16 & 0.8 & 1.32 & 0.01
    \end{array} \right)
    \end{align}

    Y$_1$ bond:
    \begin{align}
    J_{H,ij}^{\alpha\beta} = \left(\begin{array}{r|ccccc}
    &d_{xy}&d_{xz}&d_{yz}&d_{z^2}&d_{x^2-y^2}\\
    \hline
    d_{xy}& 0. & 0.08 & 0.03 & 0.2 & 0.02\\
    d_{xz}& 0.08 & 0.44 & 0.09 & 1.2 & 0.8\\
    d_{yz}& 0.03 & 0.09 & 0. & 0.07 & 0.16\\
    d_{z^2}& 0.2 & 1.2 & 0.07 & 0.7 & 1.32\\
    d_{x^2-y^2} & 0.02 & 0.8 & 0.16 & 1.32 & 0.01
    \end{array} \right)
    \end{align}

    Z$_1$ bond:
    \begin{align}
    J_{H,ij}^{\alpha\beta} = \left(\begin{array}{r|ccccc}
    &d_{xy}&d_{xz}&d_{yz}&d_{z^2}&d_{x^2-y^2}\\
    \hline
    d_{xy}& 0.4 & 0.08 & 0.08 & 0.57 & 1.37\\
    d_{xz}& 0.08 & 0. & 0.03 & 0.06 & 0.15\\
    d_{yz}& 0.08 & 0.03 & 0. & 0.06 & 0.15\\
    d_{z^2}& 0.57 & 0.06 & 0.06 & 0.29 & 0.7\\
    d_{x^2-y^2} & 1.37 & 0.15 & 0.15 & 0.7 & 1.71
    \end{array} \right)
    \end{align}

    \subsection{Biquadratic Exchange Couplings}
    
    \begin{figure}[t]
    \includegraphics[width=0.65\linewidth]{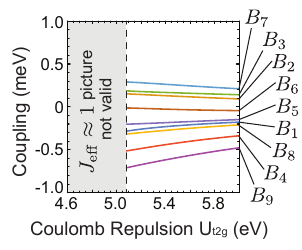}
    \caption{Intersite biquadratic interactions for Z$_1$ bond as a function of $U_{t2g}$.}
    \label{fig-biquadratic}
    \end{figure}
    
    In the limit of large $U_{t2g}$, the intersite interactions may be discussed in terms of states smoothly connected to the lowest $J_{\rm eff} = 1$ levels. In this case, the biquadratic terms may be expressed in terms of the operators:
    \begin{align}
    \mathcal{O}_i^{2,-2} =& \  S_i^xS_i^y + S_i^yS_i^x 
    \\
    \mathcal{O}_i^{2,-1} = & \ S_i^y S_i^z + S_i^z S_i^y
    \\
    \mathcal{O}^{2,0} = & \ 3S_z^2-2
    \\
    \mathcal{O}_i^{2,+1} =& \  S_i^xS_i^z + S_i^zS_i^x
    \\
    \mathcal{O}_i^{2,+2} =& \  (S_i^x)^2-(S_i^y)^2
    \end{align}
    Here, the $(x,y,z)$ axes correspond to the cubic axes defined in Fig.~1 of the main text. Defining $\mathbf{O}_i = (\mathcal{O}_i^{2,-2}, \mathcal{O}_i^{2,-1}, \mathcal{O}_i^{2,0}, \mathcal{O}_i^{2,+1}, \mathcal{O}_i^{2,+2})$, the biquadratic interactions can be written $\mathbf{O}_i\cdot \mathbb{B}_{ij} \cdot \mathbf{O}_j$. For the nearest neighbor Z-bond, the coupling matrix can be written in terms of nine parameters as:
    \begin{align}
    \mathbb{B}_{ij} = \left(\begin{array}{ccccc}
    B_1 & B_2 & B_3 & B_2 & 0 \\
    B_2 & B_4 & B_5 & B_6 & B_7 \\
    B_3 & B_5 & B_8 & B_5 & 0 \\
    B_2 & B_6 & B_5 & B_4 & -B_7 \\
    0 & B_7 & 0 & -B_7 & B_9
    \end{array}\right)
    \end{align}
    The computed values are given in Fig.~\ref{fig-biquadratic} as a function of $U_{t2g}$. It can be seen that the biquadratic couplings are indeed relatively large.

\end{document}